\documentclass[utf8]{SCNSarXiv}

\usepackage{url,hyperref,lineno,microtype,subcaption}
\usepackage[onehalfspacing]{setspace}

\def\keyFont{\fontsize{8}{11}\helveticabold }
\def\firstAuthorLast{Liang {et~al.}} 
\def\Authors{Shixiao Liang\,$^{1,*}$, Aaron Higuera\,$^{1}$, Christina Peters\,$^{2}$, Venkat Roy\,$^{3}$, Waheed~U.~Bajwa\,$^{3}$, Hagit Shatkay\,$^{2}$, and Christopher D. Tunnell\,$^{1}$}
\def\Address{$^{1}$Department of Physics and Astronomy, Rice University, Houston, TX, United States \\
$^{2}$Department of Computer and Information Sciences, University of Delaware, Newark, DE, United States\\
$^{3}$Department of Electrical and Computer Engineering, Rutgers University, Piscataway, NJ, United States}
\def\corrAuthor{Shixiao Liang}
\def\corrEmail{liangsx@rice.edu}

\begin{document}
\onecolumn
\firstpage{1}

\title[Domain-informed networks for interaction localization]{Domain-informed neural networks for interaction localization within astroparticle experiments} 

\author[\firstAuthorLast ]{\Authors} 
\address{} 
\correspondance{} 

\extraAuth{}

\maketitle

\begin{abstract}

\section{}
This work proposes a domain-informed neural network architecture for experimental particle physics, using particle interaction localization with the time-projection chamber (TPC) technology for dark matter research as an example application. A key feature of the signals generated within the TPC is that they allow localization of particle interactions through a process called reconstruction (i.e., inverse-problem regression). While multilayer perceptrons (MLPs) have emerged as a leading contender for reconstruction in TPCs, such a black-box approach does not reflect prior knowledge of the underlying scientific processes. This paper looks anew at neural network-based interaction localization and encodes prior detector knowledge, in terms of both signal characteristics and detector geometry, into the feature encoding and the output layers of a multilayer (deep) neural network. The resulting neural network, termed \emph{Domain-informed Neural Network} (DiNN), limits the receptive fields of the neurons in the initial feature encoding layers in order to account for the spatially localized nature of the signals produced within the TPC. This aspect of the DiNN, which has similarities with the emerging area of graph neural networks in that the neurons in the initial layers only connect to a handful of neurons in their succeeding layer, significantly reduces the number of parameters in the network in comparison to an MLP. In addition, in order to account for the detector geometry, the output layers of the network are modified using two geometric transformations to ensure the DiNN produces localizations within the interior of the detector. The end result is a neural network architecture that has 60\% fewer parameters than an MLP, but that still achieves similar localization performance and provides a path to future architectural developments with improved performance because of their ability to encode additional domain knowledge into the architecture.

\tiny
 \keyFont{ \section{Keywords:} astroparticle physics, direct detection dark matter, machine learning, neural network, reconstruction, time-projection chamber} 
\end{abstract}


\section{Introduction} 

Astroparticle physics has experienced a renaissance during the last decade. For instance, experiments searching for exotic phenomena related to neutrinos and dark matter particles have made significant advances in answering fundamental questions in both cosmology and particle physics~\citep{Billard:2021uyg,Giuliani:2019uno}. Often these experiments conduct extreme rare-event searches for signals that are challenging to measure, thereby imposing new requirements on the detector technology and the accompanying data-analysis methods. The focus of this paper is developing new data-analysis methods to better understand these signals and meet the scientific requirements.

Machine Learning (ML), particularly in the form of deep neural networks~\citep{lecunDL,goodfellow2016deep}, has also thrived in the last decade and enabled innovations in many fields, including experimental particle physics~\citep{Albertsson:2018maf,Radovic:2018dip}. Recent advances in deep learning concentrate on ``daily life" tasks such as computer vision and natural language processing~\citep{lecunDL,DLRev}. Methodologically, neural networks have proven adept at such problems, with numerous neural network architectures appearing that are optimized for data sets in those fields, usually inspired by the properties of the data in these fields. Research into physics-specific approaches have focused on problems closely related to theoretical physics where the measurement process is abstracted away~\citep{Komiske:2018cqr,Cranmer:2019eaq}.  
For instance, recent research in physics-informed neural networks (PINNs) focuses on solving differential equations by introducing additional terms to the loss function~\citep{2017arXiv171110561R}.
Similar work related to climate modeling shows that neural networks can emulate nonlinear dynamical systems with increased precision by introducing analytical constraints into the network architecture~\citep{PhysRevLett.126.098302}.
This paper complements and extends these prior works by developing \emph{Domain-informed Neural Networks} (DiNNs) that contain architectural constraints to encode prior knowledge of signal characteristics and detector geometry.

Our main contributions in the paper include description, implementation, and evaluation of a DiNN incorporating physics-detector domain knowledge for localization of particle interactions within a leading particle-detector technology. Section 2 describes the detector technology and the domain problem. Section 3 describes the generation process of the samples used for training a machine learning model. Section 4 explains the physics-informed neural network and describes the hard architectural constraints on its hidden and output layers. Section 5 and 6 discuss, respectively, the performance of the prototype DiNN and the impact of the new architecture in comparison to the state-of-the-art methodology. Finally, Section 7 concludes the paper.

\section{Problem of particle localization in astroparticle detectors} 

The physical properties of an interaction within a particle-physics experiment, such as the type of interaction, the amount of energy deposited, and the interaction position, are inferred using measurements from sensor arrays. The procedure for this so-called ``reconstruction" (a.k.a, inverse problem) depends on the working principle of the particle detector, while the choice of detector technology depends upon which science is being pursued.  For dark-matter direct detection, which attempt to terrestrially measure the Milky Way's dark matter wind, the most sensitive experiments use dual-phase liquid xenon time projection chambers (DP-LXeTPCs), $e.g.$ XENON~\citep{xe1t}, LUX~\citep{lux}, and PandaX~\citep{PandaXii}.

\subsection{Detection principle of a time-projection chamber}

Time-projection chambers (TPCs) are  among the most versatile particle detectors. The essential feature of TPCs is the ability to accurately measure the positions of particle interactions inside the detector. There are many successful applications of TPCs in collider experiments~\citep{startpc,alicetpc}, neutrino experiments~\citep{MicroBooNE:2016pwy,DUNE:2020txw}), as well as in the study of dark matter. DP-LXeTPCs are the world-leading detectors for dark matter direct detection \citep{Schumann:2019eaa} and are proven to be able to find extremely rare events~\citep{XENON:2019dti}. Additionally, LXeTPCs are a compelling technology for future neutrinoless double-$\beta$ decay experiments \citep{nEXO:2017nam}.

\begin{figure}
\begin{center}
\includegraphics[width=10cm]{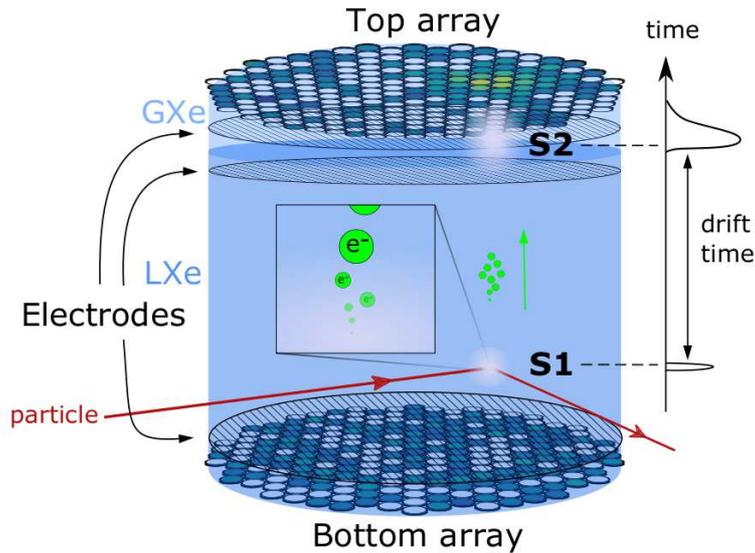}
\end{center}
\caption{ A schematic of the working principle of a DP-LXe TPC detector. Particles interact and deposit energy in liquid xenon target. An S1 signal is produced at the interaction location. Ionization electrons are released from the interaction point and drift toward the top of the TPC, where an S2 signal is produced and observed.}\label{fig1}
\end{figure}

The signal characteristics from a DP-LXeTPC have been extensively studied and are well understood.  A DP-LXeTPC consists of a liquid-xenon target observed from above and below by arrays of photosensors, where there is a buffer layer of xenon gas between the liquid and the top photosensor array as shown in Figure~\ref{fig1}. Electrodes create strong electric fields throughout the detector to drift the electrons from the target into the gaseous xenon region. When energetic particles scatter with xenon atoms in the target, the resulting recoil excites and ionizes the medium.  (This results in scintillation light, often called S1 signal, that is almost immediately observed by the photosensors.) More importantly for this study, the ionization process results in unbound electrons. These free electrons are within an electric field, so they drift toward the top of the TPC. Once the electrons reach the top of liquid xenon volume, they are extracted into the gaseous xenon by the strong electric field, causing an electron cascade that also produces scintillation light (often called S2 signal).  This S2 signal is observed by the same sensor arrays that observe the S1 signal. The 2D position, in the plane of the photosensors, of the interaction $\vec{x}_I$ is inferred from the S2 signal illumination pattern $\vec{H} \in \mathbb R_{\ge 0}^{n_t} $, where $n_t$ is the number of photosensors in the top array. In practice, only the top array is used as it is the closest to the S2.

Without loss of generality, we consider a ``generation two" (G2) TPC for direct-detection dark matter as an example application for this DiNN work.
In our example, we consider $n_t = 253$ photosensors in the top array and a detector cylinder of 67\,cm in radius. The diameter of each photosensor is assumed to be 7.6\,cm. More details on the simulation of S2 signals used for training and evaluation can be found in Section 3.

\begin{figure}
\begin{center}
\includegraphics[width=15cm]{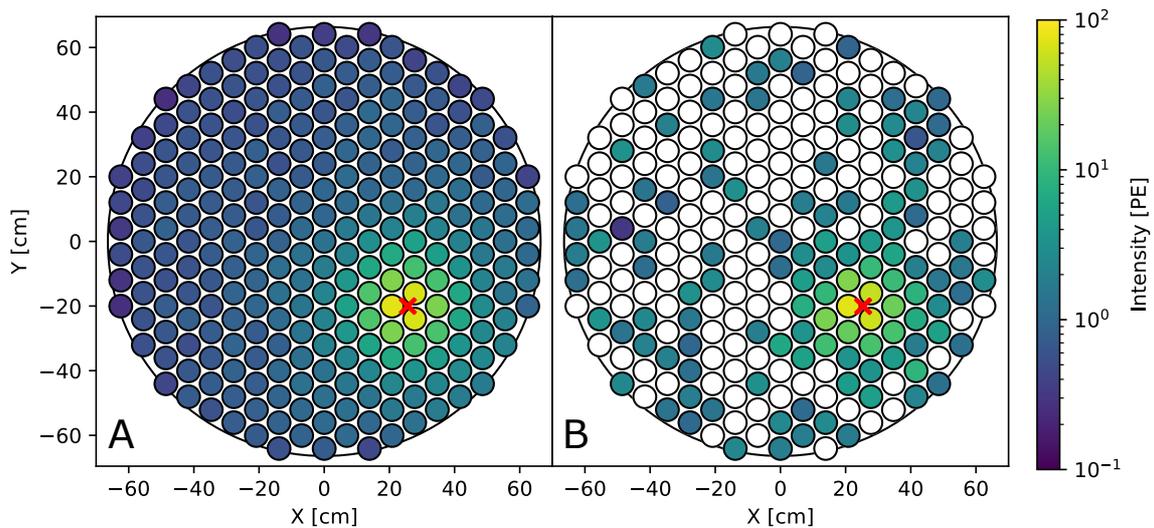}
\end{center}
\caption{\textbf{(A)}: average simulated S2 pattern. \textbf{(B)}: a single simulated S2 pattern. S2s are simulated assuming 20 electrons are extracted into the gaseous xenon at the location of the red marker. The averaged S2 pattern is ideal which allows us to infer the position perfectly, while the actual S2 pattern contains fluctuations, noises and detector effects that make accurate position reconstruction difficult.}\label{fig2} 
\end{figure}

We focus on the challenge of interaction localization as this problem is difficult and representative of most reconstruction challenges in (astro)particle physics. In our case of DP-LXeTPCs, the required uncertainty for the inferred S2 position is less than 1\,cm, one order of magnitude smaller than the length scale of the photosensors.
We are considering a rare-event search where any misreconstructed S2 positions negatively impacts the sensitivity to dark matter.
Challenges for this task include Poisson fluctuations in S2 pattern as the S2 signals are small, sometimes with only 40 photons detected across the entire array resulting from a single electron \citep{Edwards:2007nj,XENON100SE}, and detector effects such as reflection of the S2 photons off of the wall of the TPC. Figure~\ref{fig2} visually illustrates the fluctuation in the S2 pattern of a relatively small S2 signal.

\subsection{Existing techniques and limitations}

We summarize S2 localization methods developed for previous DP-LXeTPC experiments and their limitations. These methods are either used in the analyses of previous experiments or are candidates for upcoming experiments.

Likelihood fitters are used for S2 localization for DP-LXeTPCs \citep{zeplin,Simola:2018ntn,pelssers,PANDA-X:2021jua,LUX:2017lif}. While this is the most robust and accurate method in theory, it is only true if the likelihood is tractable. The likelihood is often difficult to estimate due to various detector systematics, such as the difficulty in modeling the reflectivity of the materials in the detector~\citep{levy} at the level required by the science. The reflectivity of materials also affects other methods, especially within the area close to the detector wall. Most importantly, likelihood fitters are slow compared to other methods as they often require computing the likelihood function multiple times, which makes it more difficult to be applied within a data analysis pipeline of experiments that take data at high rate.

The common method for S2 localization relies upon classical neural networks consisting of a fully-connected network, also known as multilayer perceptron (MLP) model.  Typically, such networks consist of one or two hidden layers mapping $\mathbb R_{\ge 0}^{n_t} \to \mathbb{R}^{2}$ with exponential linear unit (\textsc{Elu}) and \textsc{Linear} activation functions \citep{XENON:2019ykp, deVries}. MLP models are able to learn a simple mapping from integrated light intensity seen by a photosensor to 2D coordinates. In practice, MLP methods are significantly faster than likelihood based methods, which is important as these algorithms may operate on petabytes of experimental data with real-time rates of more than 100 MB/s. However, MLP models are the most general form of neural networks and are therefore sub-optimal for this problem. There are two major challenges in particular related to the use of MLP models for our purposes. First, since the number of trainable parameters in fully connected networks increases rather quickly as the networks get deeper~\citep{726791}, this increases the risk of overfitting due to the larger capacity of the networks~\citep{726791, Vapnik00nat}. Second, and perhaps most importantly, fully connected networks completely ignore the topology of the input data samples; the ordering of individual dimensions of data samples into the input layer can be changed without any effects on the training~\citep{726791}. Both the density of connections in MLP models and their indifference to the data topology limit our ability to interpret their outcomes.


Convolutional neural networks~(CNNs)~\citep{lecun-bengio-95a} are an example of a neural network architecture in computer vision that uses assumptions about the data topology to reduce the computational complexity of deep learning.  Specifically, convolutional layers encode the concept of ``feature locality" in the data through learning a kernel that is translational invariant (i.e. weight sharing) and can therefore e.g. detect edges and other textures.  These CNN models have successful applications in astroparticle physics experiments~\citep{NocaClassifier,NovaEnergy}, including both liquid-argon TPC~\citep{MicroBooNEcnn,MicroBooNEsparse,Grobov:2020jjx} and a liquid-xenon TPC~\citep{EXO}. However, the nature of DP-LXeTPC experimental data---and particle-physics detectors more generally---does not often lend itself to CNN techniques as the nature of this data is not image based.
Specifically, the photosensors typically are not arranged on a grid, or square lattice. It is possible to construct a contrived mapping of the photosensors array onto a square lattice for some certain arrangements~\citep{HexaConv,icecubehex}, but this geometrically distorts features in the data while also affecting the performance due to artificial empty pixels from the transformation. In addition, CNN models built with plain convolutional layers are not ideal for coordinate transform problems including localization problems~\citep{liu2018intriguing}.

\section{Simulation of S2-signal samples for training and evaluation}

Here we describe how we simulate S2 patterns based on the generation mechanism of S2 signals in DP-LXeTPC for training and evaluating the algorithms for S2 localization. The photons of one S2 signal are considered to be released from a point source at where the electrons enter the gaseous xenon and the number of photons in a S2 signal is proportional to the number of electrons entering the gaseous xenon. The number of photons recorded by photosensors obey Poisson distributions $\text{Pois}(\mu)$, where the expectation value $\mu$ is calculated by multiplying the light collection efficiency $\text{LCE}(x,y)$ of each photosensor and the number of photons in the S2 signal. The number of photons can be calculated by multiplying the number of electrons that generate the S2 signal $N_e$ by the scintillation gain $\text{SG}(x,y)$ which is the conversion factor between the number of S2 photons and the number of electrons.
The recorded S2 signal intensity of each photosensor fluctuates due to the resolution of photosensors, which is considered to obey a normal distribution $\text{Norm}(1,\sigma)$ with $\sigma$ being the resolution of photosensors. Thus the S2 intensity recorded by photosensor $i$ can be expressed as a product of a Poisson random variable and a normal random variable:

\begin{equation}\label{s2model}
    H_{i} = \text{Pois}\big( N_e \times \text{SG}(x,y) \times \text{LCE}_{i}(x,y)\big) \times \text{Norm}(1,\sigma),
\end{equation}

\noindent where $(x,y)$ is the coordinate of the true position of the S2 signal. The light collection efficiency function $\text{LCE}_{i}(x,y)$ heavily depends on the location of the photosensor and could change dramatically between photosensors.

We use a model that assumes the same light collection efficiency for all the photosensors as well as uniform scintillation gain. The light collection efficiency is modeled using an empirical function of the distance between the S2 position and the photosensor position similar to the one used in~\citep{LUX:2017lif}:

\begin{equation}\label{lce}
    \text{LCE}(\rho)= E_0 \times \frac{1-b}{\big((\rho/d)^2+1\big)^p}+ a \rho + b ,
\end{equation}

\noindent where $\rho$ is the distance between the S2 position and the photosensor position. $E_0$, $a$, $b$, $p$ and $d$ are parameters that determine the shape of this function with values shown in Table~\ref{tab:fit_lce_params} used in this work.

\begin{table}[]
    \begin{center}
    \begin{tabular}{|c|c|c|c|c|}
    \hline
      $E_0$& $a$& $b$& $p$ & $d$\\
       $1.18\times10^{-2}$ & $2.39$ & $10.3$ & $-6.77\times10^{-7}$ & $9.86\times10^{-5}$\\
       \hline
    \end{tabular}
    \caption{The model parameters used for this work for the intensity seen by an individual photosensor (Eq.~\ref{lce}) with $\rho$ in the unit of cm.}
    \label{tab:fit_lce_params}
    \end{center}
\end{table}

This light collection efficiency model does not take into account the optical effects of reflection from certain detector components. For example, the reflection from the detector wall which could largely change the light collection efficiency of photosensors placed close to the detector wall. When implementing any S2 localization algorithm trained on samples generated using the above model for an actual DP-LXeTPC detector, it is necessary to compare the model to both a data-driven and a simulation-based light collection efficiency model to ensure the appropriateness of the above approximations.

\section{Method: domain-informed neural network architecture for interaction localization}

\subsection{Encoding signal characteristics using graph-constrained hidden layers}

We encode our knowledge of the S2 signal characteristics into the network architecture through connectivity constraints. 
The photosensors closer to the S2 position $\vec{x}$ have more photons incident upon them than those farther away, and thus provide more localization information. 
Moreover, photosensors outside the S2's field-of-view might record other signals uncorrelated to the S2 signal, such as position-uncorrelated single-electron S2s \citep{XENON100SE}, which may bias the inference of the S2 location. 
Therefore, we expect that encoding this intrinsic ``locality" will be more efficient - and potentially more effective - at localizing S2s.  Our motivation is the success of CNN models in computer vision, which rely on local connections and locality as discussed above.

\begin{figure}
\begin{center}
\includegraphics[width=7cm]{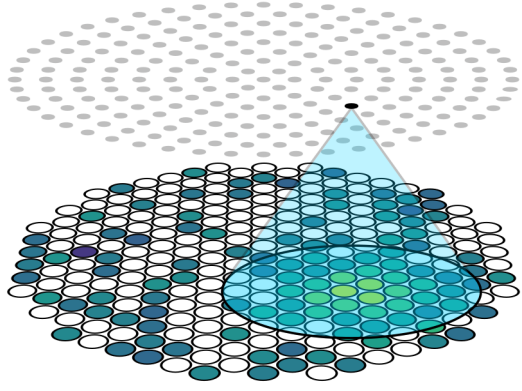}
\end{center}
\caption{ Schematic of the locality constraint. The distance between the highlighted neuron and photosensors in the blue area is smaller than a threshold. Photosensors inside the blue area are connected to the highlighted neuron while photosensors outside the blue area are not, limiting the receptive field of the highlighted neuron to be the blue area. }\label{fig:2}
\end{figure}

This is achieved by limiting the receptive field of neurons in the hidden layers. The input vector $\vec{H}_{t} \in \mathbb{R}_{\ge 0}^{n_t}$ to the network corresponds to an S2 pattern on the top sensor array, which means that there is a one-to-one relationship between photosensor positions and neurons in the input layer to the neural network model.
We introduce an architectural constraint by assigning each neuron in the following layer a 2D position, and only making edges to neurons in the following layer if they are within some distance threshold $d$ within $(x,y)$ from neurons in the input layer. In this way, every neuron only receives information from a small cluster of nearby photosensors in the receptive field of a circular disk with radius $d$ centered at the neuron's position and ignores others, as shown in Fig.~\ref{fig:2}.  This differs from the standard fully-connected technique where each neuron is connected to all neurons in the input layer. However, if the distance threshold is sufficiently large to include all neurons, the two techniques are equivalent. This technique contrasts CNNs as each neuron has its own receptive field without weight sharing as translational invariance is not guaranteed, and is similar to the idea of locally connected layers which is used in an other work in particle physics~\citep{deOliveira:2017pjk}.

In practice, the network is constructed by placing the neurons and the photosensors on the same 2D plane and constructing a graph using the rule described above, where part of the adjacency matrix of this graph is used to mask the weight matrix of the neural network layer. 
With the input to this layer represented by $\vec{H}^{l-1}\in{\mathbb R}^N$, the operation of this layer can be represented as

\begin{equation}\label{gcl}
 \vec{H}^l = f({\bf W}_l{\bf A}_l\vec{H}^{l-1} + \vec{b}_l),
\end{equation}

\noindent where ${\bf W}_l$ and $\vec{b}_l$ are the trainable weight matrix and the bias vector, respectively, $f(\cdot)$ is an activation function, and ${\bf A}_l$ is the adjacency matrix that represents the connection between the input layer and the proceeding layer.
We refer to this layer as graph constrained layer in the following sections. 

The graph constrained layer requires neurons in its input layers to have 2D positions. With neurons being interpreted as points in a 2D plane, we can extend a neural network model by adding another graph constrained layer after one. 

\subsection{Encoding detector geometry using a geometry-constrained output layer}

We describe the design of a hard detector-inspired constraint on the output space for the neural network model. This constraint enforces the model to produce physically meaningful predictions located only within the detector.

We must modify the output layer of the network to incorporate this constraint.
Historically, the localization regression is performed using two neurons with linear activation functions at the output layer since we are predicting a 2D position. The possible output space for these models is $\mathbb{R}^2$.  
Meanwhile, S2s can only be produced inside the detector, which means the physically meaningful output space is a circular disk $D(R_{TPC}) = \left\{ \vec{x} \in \mathbb{R}^2 \,:\left\| \vec{x} \right\| _2 =  \sqrt{|x|^2 + |y|^2} < R_{TPC} \right\}$.  It is important to require that the predicted positions are within the detector $D(R_{TPC})$ as there are position-dependent corrections for each interaction that otherwise are ill-defined.  Furthermore, due to the self-shielding properties of liquid xenon, most S2 signals are near the edge of the detector and are at higher risk of being localized outside the detector.  This effect is also more significant for smaller S2s as they are harder to localize.

\begin{figure}
\begin{center}
\includegraphics[width=10cm]{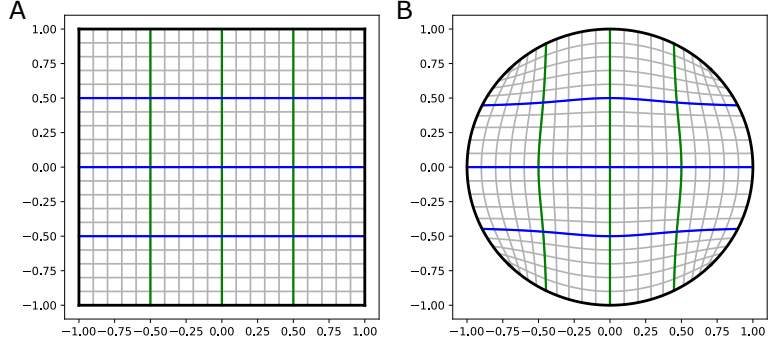}
\end{center}
\caption{ The analytical mapping from square to circle. \textbf{(A)}: a square in 2D space $T = \left\{ \vec{x} \in \mathbb{R}^2 \,: |x| < 1, |y| < 1 \right\}$. \textbf{(B)}: unit circle, mapped from the square using Eq.~\ref{eq:01}. The lines in the square are mapped onto the curves in the circle.}\label{fig:3} 
\end{figure}

The first part of this hard constraint is constraining the output space to be the limited space instead of the infinite $\mathbb{R}^2$ space. We use the \textsc{Tanh} activation function in the output layer to make a temporary output in a 2D square  $T = \left\{ \vec{x} \in \mathbb{R}^2 \,: |x| < 1, |y| < 1 \right\}$.  The second part is analytically mapping the temporary \emph{square} $T$ into a unit \emph{circle} $D(1)$ using Fernandez Guasti \emph{squircle} (FG-squircle) mapping, as shown in Fig.~\ref{fig:3}.~\citep{LambersMappings,fong, FGnotes} and scale it to the radius of the TPC $R_{TPC}$. The square to disk mapping is shown in Eq.~\ref{eq:01}.

\begin{equation}
    x' = R_{TPC}\frac{x\sqrt{x^2 + y^2 - x^{2}y^{2}}}{\sqrt{x^2 + y^2}},\quad
    y' = R_{TPC}\frac{y\sqrt{x^2 + y^2 - x^{2}y^{2}}}{\sqrt{x^2 + y^2}}
    \label{eq:01}
\end{equation}

\subsection{Implementation}

\begin{figure}
\begin{center}
\includegraphics[width=17cm]{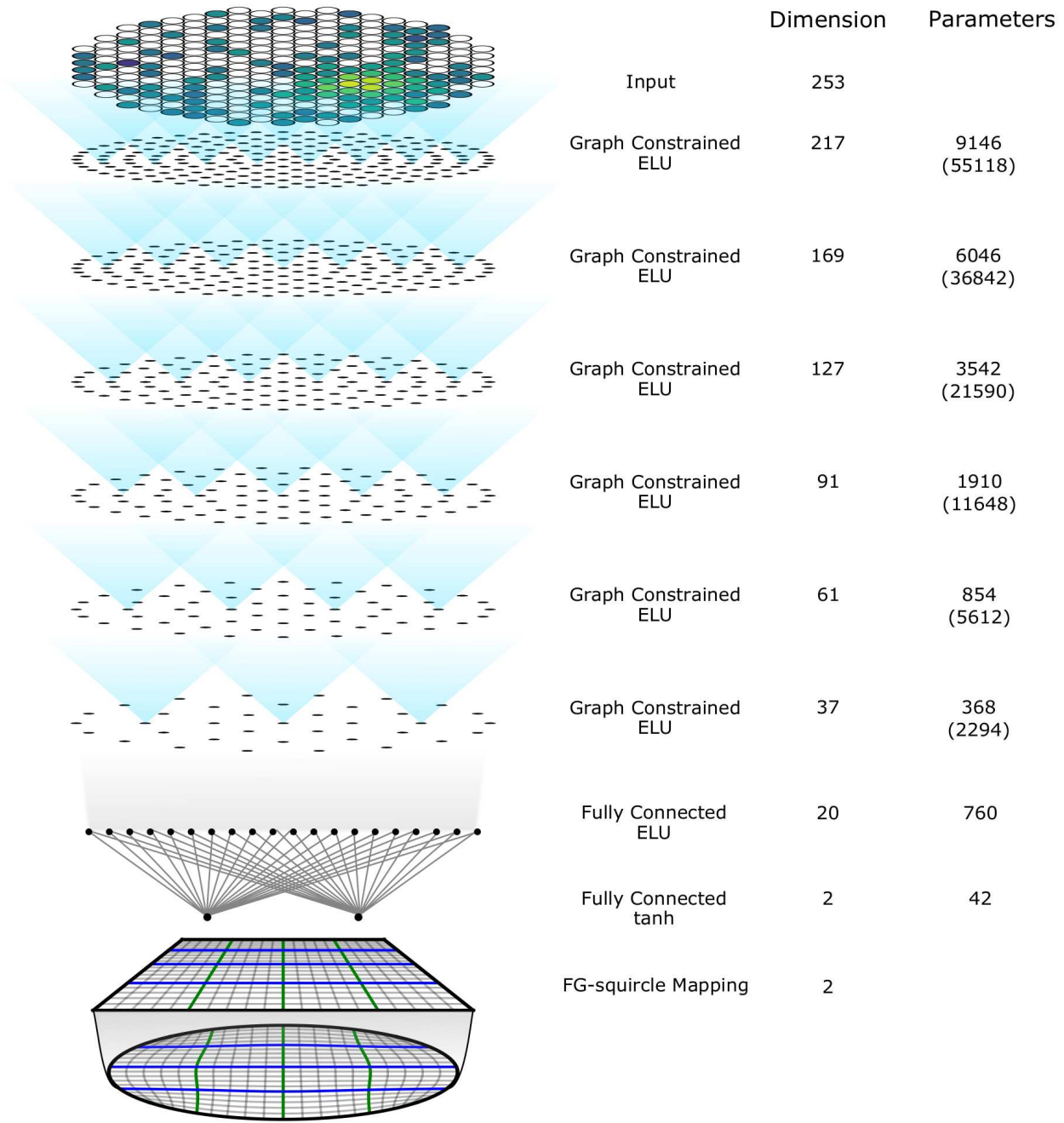}
\end{center}
\caption{The architecture of the implemented prototype model. The dots in graph constrained layers represents the assigned positions of the neurons. The numbers in brackets are the number of parameters if the neurons are fully connected for graph constrained layers.}\label{fig:4} 
\end{figure}

We implemented a prototype neural network model using the graph-constrained layer and the constraint on the output space. The architecture of this neural network model is shown in Figure~\ref{fig:4}. The input vector $\vec{H}_{t} \in \mathbb{R}_{\ge 0}^{n_t}$ is embedded into an input layer with dimension 253 alone with the positions of photosensors. After the input layer there are 6 graph-constrained layers, with dimension [217, 169, 127, 91, 61, 37] respectively. In each layer, the assigned positions for the neurons are arranged in rings and distributed roughly uniformly in a circle that is approximately the same area as the TPC cross section. The threshold for connection used for this model is 30\,cm and justification for this choice is discussed in Section~\ref{Optimization}. There are two fully connected layers after the graph-constrained layers, with the last one using \textsc{Tanh} squashing function to produce a temporary output constrained to a filled square. The FG-squircle mapping is implemented to map the temporary square output onto the desired disk output space. The number of trainable parameters in this model is 22,921, less than 0.2 times the number of parameters in a fully connected network of the same architecture as this model.

The model is implemented with TensorFlow~\citep{TF}. We train the model on $5\times 10^5$ simulated S2 patterns for 250 epochs with Adam optimizer~\citep{adam} using mean squared error as loss function. A adaptive learning rate scheduler is applied to lower the learning rate when the validation loss hits a plateau for more than 10 epochs. This model is referred to as a DiNN model in the following sections.

\section{Evaluating performance of interaction localization}\label{eva}

Here we define the metrics used to evaluate the performance of methods developed for S2 localization. The models are tested on simulated S2 signals with known true positions, which is needed for the calculation of these metrics. We then compare the performance of the DiNN model and a 4-layer MLP model trained on the same data set with the same optimizer and training configuration using these metrics. The MLP model, which is very similar to the one used in~\citep{XENON:2019ykp}, has 57,842 trainable parameters.

We use $\vec{x}$ to denote the true S2 position used in simulation and $\hat{\vec{x}}$ to denote the prediction based on the model for a given S2 pattern. $R = |\vec{x}|$ is the distance of the true S2 position from the center of the detector, commonly called the radius of S2. The difference between the true and predicted S2 positions, $\hat{\vec{x}} - \vec{x}$, is the localization error. As mentioned in section 3, the S2 pattern is dependent on the number of electrons that generate the S2 signal. We mainly use simulated S2s generated by 100 electrons, referred to as 100 electron S2s, because this is approximately the intensity of S2 signals caused by dark matter particle interactions.

\begin{figure}
\begin{center}
\includegraphics{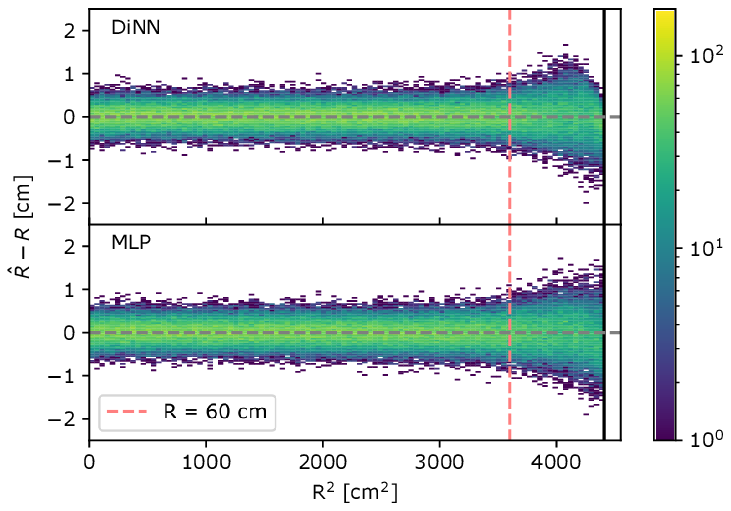}
\end{center}
\caption{Radial error $\hat{R} - R$ on 100 electrons simulated S2s. A positive radial error means the S2 is localized further away from the center of the detector than the true position and vice versa. Both the DiNN and MLP model cannot localize S2s in $R>60$~cm region as accurate as S2s in $R<60$~cm region, resulting in a wider spread in $R>60$~cm region and the positive radial error is suppressed by the constrained output in the region extremely close to the detector wall for the DiNN.} \label{fig:rerr} 
\end{figure}

When searching for rare events, such as dark matter particle interactions, the radius of an interaction is more commonly used than the $(x,y)$ coordinate in statistical inference~\citep{XENON:2019ykp}. Figure~\ref{fig:rerr} shows the distribution of the difference between the true and predicted S2 radii, $\hat{R} - R$, on simulated 100 electron S2s as a function of the true radius, $R$, for the MLP and DiNN model. Notice in the $R < 60\ $\,cm region, the S2s are localized with radius error less than 1\,cm, and $\hat{R} - R$ has no dependence on true radius, which is compatible with the goal of setting the uncertainty of S2 localization to be one order of magnitude lower than the scale of the photosensors. In the $R > 60\ $ cm region, both the GCN and MLP models cannot maintain the same performance as they do in $R < 60\ $\,cm region. The deteriorated performance could result from incomplete sampling of S2 signals in this region since there are no photosensors outside the detector. 
The large difference in the behavior in the $R < 60\ $ cm and $R > 60\ $\,cm regions by both the DiNN and MLP models motivates a comparison of the two regions using other metrics.

\begin{figure}
\begin{center}
\includegraphics[width=15cm]{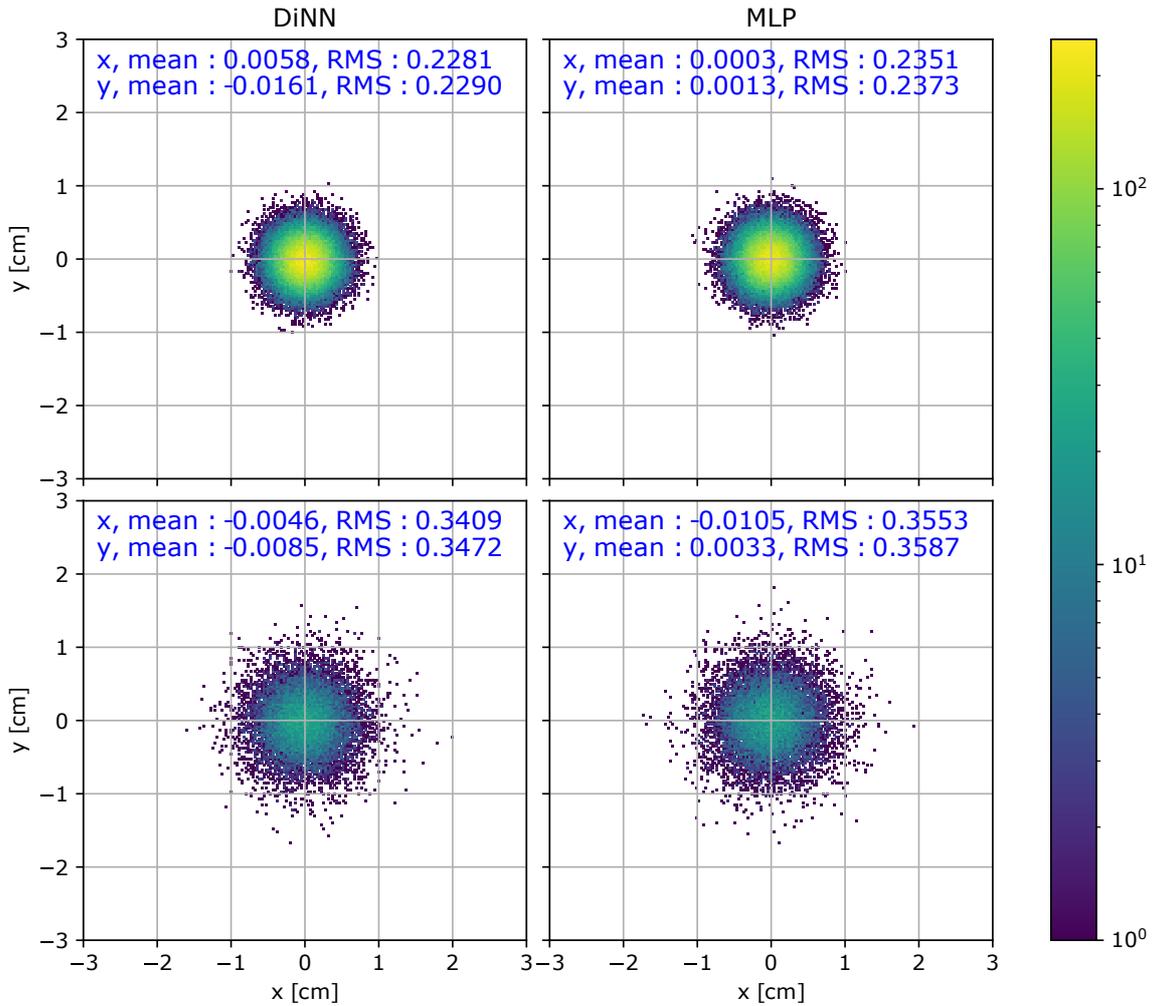}
\end{center}
\caption{Distribution of localization error, $\hat{\vec{x}} - \vec{x}$, for a test set of simulated 100 electron S2 signals. S2 signals with $R < 60 cm$ (top row) have a smaller RMS than interactions with $R > 60 cm$ (bottom row).}\label{fig:err} 
\end{figure}

While the $(x,y)$ coordinate is not directly used for physical analysis, it is useful for diagnostic analysis for the detector. A high position resolution is crucial for identifying the topology of interactions~\citep{Wittweg:2020fak}. Thus the distribution of localization error, $\hat{\vec{x}} - \vec{x}$, is useful for evaluating the performance of the models. The distribution of localization errors is shown in Figure~\ref{fig:err}. We calculate the mean value and the root mean squared (RMS) of the localization error for quantitative comparison. A small mean value in both the $x$ and $y$ direction indicates that the localization is not biased towards any direction, which appears to be the case for both the DiNN and MLP model in both $R < 60\ $ cm region and $R > 60\ $ cm region. A smaller RMS indicates more accurate localization. The DiNN model has similar RMS in both regions to the MLP model.

\begin{figure}
\begin{center}
\includegraphics[width=15cm]{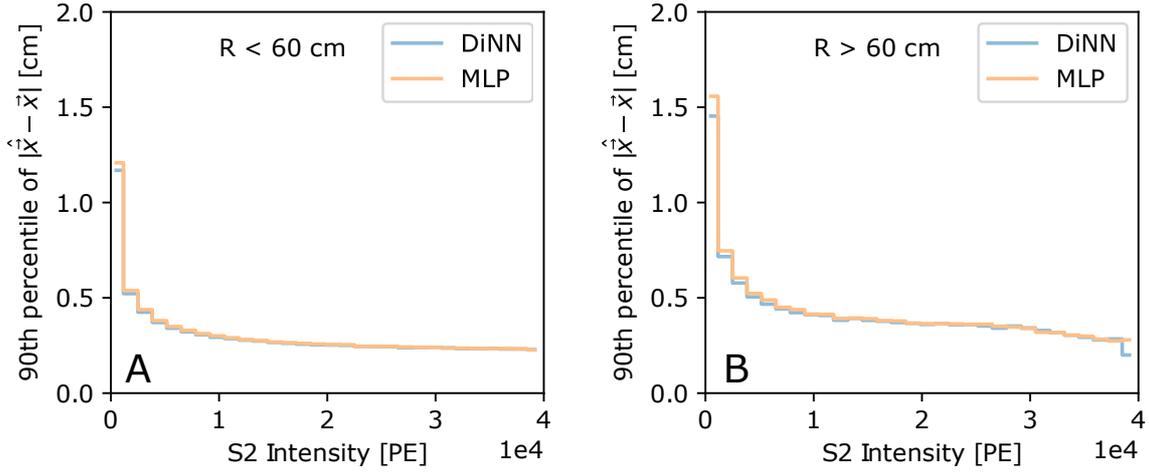}
\end{center}
\caption{ The 90th percentile of localization error as a function of S2 intensity, for \textbf{(A)}: $R < 60\ $ cm and \textbf{(B)}: $R > 60\ $ cm. S2s with higher intensities are localized with smaller error by both models, and S2s at $R > 60\ $ cm region are localized with larger error than S2s at $R < 60\ $ cm region, both as expected.}\label{fig:90th} 
\end{figure}

The difficulty of localizing S2s varies with the intensity of S2. The 90th percentile of localization error can provide us an intuitive understanding of the resolution of localization as a function of S2 intensity level. As shown in Figure \ref{fig:90th}, the 90th percentile of localization error drops for both the DiNN and MLP model as S2 intensity increases. Again the 90th percentile of localization error is much larger in $R > 60 \ $\,cm region. In both regions the DiNN model has marginally smaller 90th percentile of localization error than the MLP model, which is consistent with the result from 100 electrons S2s shown in Figure \ref{fig:err}.

\begin{figure}
\begin{center}
\includegraphics[width=8cm]{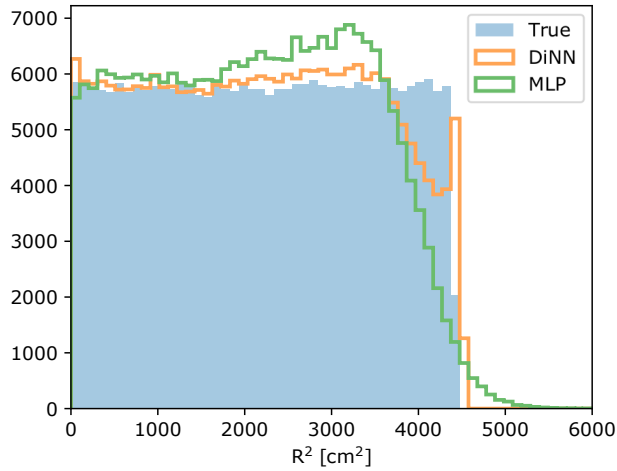}
\end{center}
\caption{Radial distribution of simulated single electron S2 signals. The MLP model localizes 2,346 of 250,000 S2s outside the detector, while the DiNN model localizes all of them inside the detector with the output constraint.}\label{fig:SE} 
\end{figure}

As stated in Section 4, smaller S2s are more sensitive to being localized outside the detector by the MLP model. We tested both the GCN and MLP model on simulated single electron S2s, which are the smallest possible S2s in the DP-LXeTPC detectors, to demonstrate the effect and determine how it is changed by the constrained output. Figure~\ref{fig:SE} shows the result of this test. The MLP model localizes a considerable amount of single electron S2s outside the detector, while the DiNN model localizes all the S2s inside the detector.

\section{Optimizing the tunable parameters in the method}\label{Optimization}

We introduced a parameter (distance threshold) in the graph constrained layer to determine its connections to the input layer. A smaller threshold means sparser connections and weaker representation power. A larger threshold means denser connections and stronger representation power, but also more parameters in the layer. When the threshold is large enough, the layer becomes fully connected, which is identical to the basic building block of MLP models. Intuitively, a threshold roughly equal to the scale of the region where a photosensor can receive light is the most meaningful and thus should work well. A well selected threshold is essential for models built with graph constrained layers to achieve maximum performance with the least parameters.

\begin{figure}
\begin{center}
\includegraphics[width=10cm]{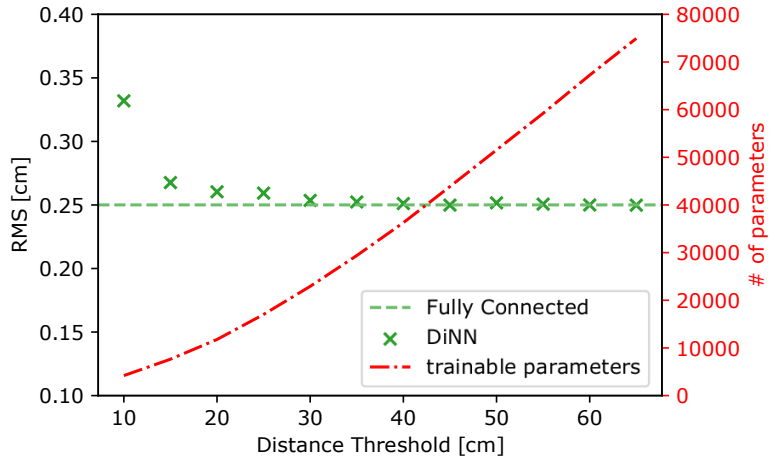}
\end{center}
\caption{The RMS of localization error (left axis) and the number of trainable parameters in the DiNN model (right axis) as functions of threshold used for the graph constrained layers in the model. The RMS decreases as the threshold increases until the threshold reaches 30 cm, but the number of parameter still increases as the threshold increases. The RMS is calculated on models' prediction of simulated 100 electron S2s and averaged over x,y dimension. }\label{fig:opt}
\end{figure}

To find the optimal threshold for the DiNN model, we trained a series of models of the same architecture, but with different distance thresholds. We also trained a DiNN model with a sufficiently large threshold so that all the graph constrained layers are fully connected. The purpose of this model is to set the limit for the performance that can be reached by the DiNN model. The total number of trainable parameters in the fully connected DiNN model is 133,906. All the models are evaluated on a test set of simulated 100 electrons S2s and the result is shown in Figure \ref{fig:opt}. When the threshold is small, the network does not have enough parameters to represent the mapping from S2 patterns to S2 positions. Increasing the threshold lowers the RMS of the distribution of localization errors. However, when the threshold is larger than 30 cm, the RMS reaches the same level as that using the fully connected DiNN model. Further increasing threshold does not lower the RMS of the distribution, but still increases the number of parameter in the DiNN model. In conclusion, a threshold of 30~cm appears to be optimal and thus is used for building the prototype DiNN model.

The two components of the DiNN model, the main body built with graph constrained layer and the geometry-constrained output layer are not closely related and could be applied to models separately. Particularly, the \textsc{Tanh} activation function and the FG-squircle mapping used for the constrained output layer introduces nonlinearity into the model, which might create difficulty for the model to learn the mapping from S2 patterns to S2 positions, especially at the regions close to the detector wall. To evaluate the effect of the main body built with graph constrained layer and the geometry-constrained output layer, we build and train MLP models and DiNN models with both linear output and geometry-constrained output and compare the RMS of the distribution of the localization errors on 100 electrons simulated S2s. The results are shown in Table~\ref{tab:ablation}. The models with constrained output have similar RMS in both $R < 60\ $ cm region and $R > 60\ $ cm region, which indicates that the nonlinearity introduced by the constrained output is not too complex for neural network models to learn and is not making localization at the region close to the detector wall difficult. 

\begin{table}[]
    \begin{center}
    \begin{tabular}{c c|c c}
    \hline
      Architecture & Output & RMS [cm] ($R<60\,$ cm) & RMS [cm] ($R>60\,$ cm)\\
     \hline
       4-layer MLP & Linear & $0.2363$ & $0.3558$\\
       4-layer MLP & Constrained & $0.2331$ & $0.3406$\\
       DiNN & Linear & $0.2314$ & $0.3470$\\
       DiNN & Constrained & $0.2280$ & $0.3419$\\
       \hline
    \end{tabular}
    \caption{Results for different combinations of neural network architectures and output layers. 4-layer MLP with linear output is the same as the MLP model discussed in Section~\ref{eva} and DiNN with constrained output is the same as the model shown in Figure~\ref{fig:4}. The RMS is calculated on models' prediction of simulated 100 electron S2s and averaged over x,y dimension.}
    \label{tab:ablation}
    \end{center}
\end{table}

\section{Conclusion and future work}

We introduce the concept of Domain-informed Neural Networks (DiNNs) for the application of the localization of S2 signals in experimental astroparticle physics. Using our prior knowledge of the S2 signal characteristics, we introduce an architectural constraint that limits the receptive field of the hidden layers. Using our prior knowledge of the detector geometry, we additionally introduce a constraint and geometrical transformation on the output layer.

A prototype DiNN model built with these two constraints is trained and tested on S2s simulated from a generic G2 dark matter search. This prototype model reached the same level of performance as a multilayer perceptron (MLP) model while containing 60\% fewer trainable parameters. 
Therefore, with a careful selection of the distance threshold related to how far a sensor `sees', the graph-constrained layers within the DiNN can greatly reduce the number of trainable parameters without degreding the performance of the prototype DiNN model. Additionally, the network is more interpretable in the sense that the output constraint puts meaningful and practical limit on the outputs of this regression problem for the first time. Such a constraint can also be used for other neural network models for regression problems that predict positions within a detector using sensors with a limited field of view.
The physics-informed locality constraint can further be applied to other astroparticle detectors such as liquid argon TPCs. Additionally, the method used for constructing this constraint can be transferred to other problems with irregular sensor arrangements. 

The idea behind graph-constrained layers has some similarities with attention-based methods, which have proven successful in many tasks including localization-related computer vision task~\citep{carion2020end} as they focus on parts of the inputs. The attention mechanism allows neural networks to focus on parts of the input in a way that is learnt from data, but requires more computational resources. We are interested in efficiently combining our method with features that are learned in a data-driven manner by attention-based methods.
Our future work is aimed at using these models on the spatiotemporal data, with a focus on calorimetry or signal detection.




\section*{Funding}

We acknowledge support from the National Science Foundation through awards 1940074, 1940209, and 1940080.  We thank NVIDIA for their support by providing GPUs.

\section*{Acknowledgments}
We thank Luis Sanchez and Alejandro Oranday for helpful discussions at the early stage of this work.


\section*{Data Availability Statement}

The code to reproduce this study can be found in \cite{liang_s_2021_5771941} or \url{https://github.com/DidactsOrg/DiNN}.

\bibliographystyle{frontiersinSCNS_ENG_HUMS} 
\bibliography{ref}

\end{document}